\RequirePackage{fix-cm}
\documentclass[smallextended]{svjour3}       
\smartqed  
\usepackage{graphicx}
\usepackage{float}
\usepackage{caption}
\usepackage{algorithm}
\usepackage{algorithmicx}
\usepackage{algpseudocode}
\usepackage{amsmath}
\usepackage{graphics}
\usepackage{epsfig}
\usepackage{bbding}
\begin{document}

\title{An objective-adaptive refinement criterion based on modified ridge extraction method for finite-time Lyapunov exponent (FTLE) calculation
}


\author{Haotian Hang         \and
        Bin Yu               \and
        Yang Xiang         \and
        Bin Zhang\footnote{Corresponding author.}             \and
        Hong Liu
}


\institute{B.Zhang(\Envelope)\at School of Aeronautics and Astronautics, Shanghai Jiao Tong University
               \\
              Tel.: +8613817855406\\
              \email{zhangbin1983@sjtu.edu.cn}           
}

\date{Received: date / Accepted: date}

\maketitle

\begin{abstract}


High-accuracy and high-efficiency finite-time Lyapunov exponent (FTLE) calculation method has long been a research hot point, and adaptive refinement method is a kind of method in this field.
The proposed objective-adaptive refinement (OAR) criterion can put adaptive particles at the vicinity of FTLE ridges.
The FTLE ridge is extracted  not simply by the magnitude of FTLE (would cause false negative refinement) and error (would cause false positive refinement) but by a modified gradient climbing method.
Moreover, the refinement regions converge to some certain region because of the objectivity of the refinement region.
Testing cases include Bickley jet, mild FTLE ridge, and experimental single vortex. The results demonstrate that the proposed algorithm can avoid useless refinement of other methods in some certain areas, and thus reduce error up to 25\% compared with the other methods with little difference in the calculation burden.
\keywords{flow structures visualization \and Lagrangian Coherent Structures (LCS) \and Objective-Adaptive Refinement (OAR) \and Finite time Lyapunov exponent (FTLE)}
\end{abstract}
\section{Introduction}
\label{intro}
Visualization of the flow field has always been one of the hot topics in the study of fluid dynamics. Details of complex unsteady flows, such as turbulent flow, vortices and shear layer is hard to demonstrate only by its velocity field and vorticity field. Thus, using post-treatment methods is important to visualize the flow field more clearly \cite{Haller2015Lagrangian}.

Researches have proposed numerous flow structure extraction methods based on Euler and Lagrangian considerations for the computational visualization of a flow field. The method based on Euler consideration, such as Q-criterion \cite{Hunt1988Eddies}, $\Delta$-criterion \cite{Chong1990A}, ${\lambda _2}$-criterion \cite{Jeong1995On}, which can give a good visualization on transient phenomenon.
In 2000, Haller \emph{et al.} \cite{Haller2000Lagrangian} systematically studied the Lagrangian coherent structures (LCSs) to describe the most attracting, repelling, and shearing material lines in a flow field. LCSs have long been used widely applied in the visualization of complex flow field \cite{Tang2010Accurate,Shadden2007Transport,Beron2015Dissipative,Karch2016Visualization,Tallapragada2011Lagrangian}.

In order to find material interface in flow field, Haller \emph{et al.} \cite{Haller2001Distinguished} demonstrated that the extremum of finite-time Lyapunov exponent (FTLE) can be seen as LCS ridge. Finding these surfaces from experimental or numerical results can give an easy but useful way for us to have an understanding of the material transport and mixture in the flow field \cite{Haller2015Lagrangian}. To illustrate the flow filed even more clearly and accurately, elliptic LCSs \cite{Blazevski2014Hyperbolic}, parabolic LCSs \cite{Froyland2014Almost} and hyperbolic LCSs \cite{LUU2008HYPERBOLICITY} are given. Parabolic LCSs are unsteady zonal jet cores in the flow field \cite{Hadjighasem2014Geodesic}, and the evolving parabolic LCS can act as a transport barrier which can prevent the mixing across the coherent structure. Elliptical LCSs are the Lagrangian vortex boundaries, which can identify the boundary of vortex rings \cite{Karrasch2015Automated}. Hyperbolic LCSs are repelling and attracting LCSs, which can converge an area to a single line. In most cases, hyperbolic LCSs coincides with the FTLE ridges, however, in some particular cases in real life, they are not the same \cite{Haller2011A}.
Although these calculation are mainly based on the eigenvector of the Cauchy-Green tensor, FTLE, as the eigenvalue of the  Cauchy-Green tensor, can also express the characteristic of the tensor. Thus, developing new methods to promote the efficiency of the FTLE calculation is still important.

%

Adaptive mesh refinement (AMR) is a method for adapting the accuracy of a solution within certain sensitive or turbulent regions of simulation \cite{Plewa2005Adaptive,Huang2011Adaptive}. This technique has been widely applied in CFD and proved to reduce the cost of calculation \cite{Wang2016A,Ng2007Adaptive,Jimbo2003Numerical,Keith1991Modeling,Berger1989Colella}. AMR has also been used in FTLE computation, which can enhance the efficiency of calculation by a large amount.
Garth \emph{et al.} \cite{Garth2007Efficient} first developed the method of AMR in LCS in 2007. Miron \emph{et al.}  \cite{Miron2012Anisotropic} applied AMR in FTLE calculation in anisotropic mesh using a commercial software. Based on this, Fortin \emph{et al.} \cite{Fortin2015A} compared various kinds of adaptive methods to provide a more accurate method. The refinement criterion of all these methods are based on the error of FTLE field.


However, the AMR in LCS methods based on error may have many problems, and fail to find FTLE ridge accurately in cases mentioned in Sec.\ref{sec:4.2}. For this reason, an adaptive refinement method based on the extraction of FTLE ridge should be proposed. Sadlo. F  \emph{et al.} \cite{Sadlo2007Efficient} applied the method based on the magnitude of FTLE \cite{Lipinski2010A} into the adaptive refinement of FTLE calculation. However, as FTLE ridge is not definitely the maximum of FTLE value, especially  when the FTLE field is undulate because of the error in initial data. A better ridge extraction method provided in \cite{Mathur2007Uncovering} is aimed as finding FTLE ridge in chaotic flow field. However, it failed to find FTLE ridge when the FTLE ridge is relatively mild as illustrated in Sec.\ref{sec:4.1.2}.
In this study, an objective-adaptive refinement (OAR) criterion based on the unique feature of FTLE ridge is proposed. According to the physical feature of FTLE ridge, a new refinement criterion is provided. This criterion can find FTLE ridge and refine the calculation efficiently.

The rest of the paper is organized as follows. In Sec.\ref{sec:2}, the principle and definition of LCS is introduced. In Sec.\ref{sec:3}, the OAR method is described and applied in FTLE calculation. Two analytical examples and one experimental example are demonstrated in Sec.\ref{sec:4}. The result and visualization obtained by different methods are also compared. The conclusions are presented in Sec.\ref{sec:6}.

\section{Finite-time Lyapunov exponent (FTLE) description}
\label{sec:2}
\subsection{The definition of FTLE}
\label{sec:2.1}
Transportation in dynamical system is often researched by the trajectory of spatial particles. Lyapunov exponent can be used in the moving trajectory of spatial particles. Finite-time Lyapunov exponent (FTLE) marks as $\sigma _t^T\left( {\bf{x}} \right)$, is a scalar quantity. When used in the computation of fluid, FTLE expresses the average separation degree for a fluid particle and the trajectory of its surrounding fluid particles from $t$ to $t+T$. The computational formula of $\sigma _t^T\left( {\bf{x}} \right)$ is as follows:
\begin{equation}\label{eq:LCS1}
\sigma _t^T\left( {\bf{x}} \right) = \frac{1}{{\left| T \right|}}ln\sqrt {{\lambda _{\max }}\left( \Delta  \right)},
\end{equation}
where \(\Delta \) is a symmetrical matrix and the result of the moving trail spatial derivative matrix multiplies its transposition.\({\lambda _{\max }}\left( \Delta  \right)\) is its maximum eigenvalue of $\Delta$. The moving trail of the fluid particle from $t$ to $t+T$, is marked as $\phi _t^{t + T}\left( {\vec x} \right)$. Then the symmetrical matrix \(\Delta \) is as follows:
\begin{equation}\label{eq:LCS2}
\Delta {\rm{ = }}{\left[ {\frac{{d\phi _t^{t + T}\left( {\bf{x}} \right)}}{{d{\bf{x}}}}} \right]^*}\frac{{d\phi _t^{t + T}\left( {\bf{x}} \right)}}{{d{\bf{x}}}}
\end{equation}
where ${\frac{{d\phi _t^{t + T}\left( {\bf{x}} \right)}}{{d{\bf{x}}}}}$ denotes the Jacobian of ${\phi _t^{t + T}\left( {\bf{x}} \right)}$, and $*$ refers to matrix transposition.

\subsection{Literature review of fast FTLE calculation}
\label{sec:2.2}
\subsubsection{Adaptive mesh refinement(AMR) based on error used in FTLE calculation}
\label{sec:2.2.1}
There are various studies focusing on the adaptive mesh refinement used in FTLE calculation\cite{Lipinski2010A,Miron2012Anisotropic,Garth2007Efficient}.
Through the definition of error is different, the inherent property is the same.
In this paper, the error in Sec.\ref{sec:4} is defined as in \cite{Garth2007Efficient}. This error can be described as follows:
\begin{equation}\label{eq:error}
\begin{array}{l}
\varepsilon  = \frac{{\sum\limits_{i = 1}^{{N_{particles}}} {{\varepsilon _i}} }}{{{N_{particles}}}}\\
{\varepsilon _i} = \left\| {S{f_i} - {f_i}} \right\|
\end{array},
\end{equation}
where ${S{f_i}}$ is the interpolation of ${f_i}$ by the surrounding particles.

\subsubsection{Ridge extraction method}
\label{sec:2.2.2}
Extracting FTLE ridge in the data has long been a research hot point \cite{Mathur2007Uncovering}. Applying this method into the adaptive refinement of FTLE calculation has began in \cite{Sadlo2007Efficient}. \cite{Sadlo2007Efficient} uses the magnitude of FTLE as the criterion of refinement as in Equ.\ref{eq:FTLErefinement}.

\begin{equation}
\label{eq:FTLErefinement}
\sigma _t^T{\left( {\bf{x}} \right)_{local}} \ge \alpha \sigma _t^T{\left( {\bf{x}} \right)_{\max }}
\end{equation}

This method may give incorrect refinement region when there are noise in the origin data.
\cite{Mathur2007Uncovering} provides us a new method to find accurate ridge in noisy data. The process is as follows. Firstly, the mesh nodes were chosen using Equ.\ref{eq:prlrefinement}.

\begin{equation}
\label{eq:prlrefinement}
{\left| {\nabla \sigma _t^T({\bf{x}})} \right|_{local}} \ge \alpha {\left| {\nabla \sigma _t^T({\bf{x}})} \right|_{\max }}
\end{equation}

Then, the particles are moved through vector $\nabla \sigma _t^T({\bf{x}})$. The cease conditions are shown in Equ.\ref{eq:conditiona} and Equ.\ref{eq:conditionb}.

\begin {equation}\label{eq:conditiona}
\min \{ {\lambda _{1,2}}({\nabla ^2}\sigma _t^T({\bf{x}}))\}  < 0
\end{equation}

\begin{equation}\label{eq:conditionb}
\theta \left\langle {e_t^T{{({\bf{x}})}_{\;{\nabla ^2}\sigma _t^T({\bf{x}})}},\nabla \sigma _t^T({\bf{x}})} \right\rangle \; \le {\theta _{\min }}
\end{equation}

However, these methods have problems and fail to refine right region in cases, such as in Sec.\ref{sec:4.1.2} and \ref{sec:4.2}.
Our method introduced in Sec.\ref{sec:3} is based on this method, but the cease condition has been revised.
\section{Proposed OAR based on ridge extraction method}
\label{sec:3}

Nonadaptive LCS calculation consumes considerable computational resources. If more resources are allocated to the vicinity of LCS ridges and less attention is paid on relatively smooth regions, adaptive methods will save significant computational resources. The proposed whole algorithm is shown as Alg.(\ref{algorithm}) in Sec.\ref{sec:3.3}.

Several AMR methods in LCS use the traditional AMR method, which focuses on error (\cite{Fortin2015A,Garth2007Efficient}) or the magnitude of FTLE (\cite{Lipinski2010A,Sadlo2007Efficient}). In this article, the proposed OAR criterion put refinement particles at the vicinity of LCS, which is discussed detailedly in Sec.\ref{sec:3.1}.

\subsection{Proposed Whole Algorithm}
\label{sec:3.3}
In this paper, our method includes two layers of refinement particles, which can be illustrated as Alg.(\ref{algorithm}). From line 2 to line 8 in Alg.(\ref{algorithm}), the level-1 particles are dispersed as a cartesian coordinate in a form of one-dimension index, and FTLE of them are calculated just like the traditional LCS method. After the calculation, the level-2 particles are dispersed as mentioned in Sec.\ref{sec:3.1}. The index of the level-2 particles is linked with the level-1 particles. From line 9 to line 19, the particles are chosen based on an appropriate $\alpha$, and then the FTLE field of the level-2 particles are calculated by the APs through particle motion and FTLE calculation. Then, all the results are put into a new mesh for the next step calculation. The dispersion and calculation of level-3 particles is similar to that the one of the level-2 particles except for the refinement rule is based on FTLE as shown from line 20 to line 29 in Alg.(\ref{algorithm}).
\begin{algorithm}[H]
\caption{Proposed OAR for LCS}
\label{algorithm}
     \begin{algorithmic}[1]
        \State read data

        \Comment {the first layer}

        \State initialize the level-1 particles
        \For {time \( < \) endtime}
        \State	interpolate the velocity in the flow field(for theoretical examples, calculate the velocity directly from time and position)
        \State  let the particles move during $\Delta t$ by the velocity
        \State	time$ \leftarrow $time + $\Delta t$
        \EndFor
        \State	calculate FTLE of each particle

        \Comment {the second layer}

        \State find the appropriate $\alpha$
        \State	find the level-1 particles which need refinement
        \State	displace the level-2 particles
        \State  link the index of the level-2 particles with the index of the level-1 particles
        \For {time \( < \) endtime}
        \State	interpolate the velocity of particles in the flow field(for theoretical examples, calculate the velocity directly from time and position)
        \State	let the particles move during $\Delta t$ by the velocity
        \State	time$ \leftarrow $time + $\Delta t$
        \EndFor
        \State	calculate FTLE of each particle
        \State	put the result into a new mesh using index linking

        \Comment {the third layer}

        \State	find the level-2 particles which need refinement
        \State	displace level-3 particles
        \State  link the index of the level-3 particles with the index of the level-2 particles
        \For {time \( < \) endtime}
        \State	interpolate the velocity of particles in the flow field(for theoretical examples, calculate the velocity directly from time and position)
        \State	let the particles move during $\Delta t$ by the velocity
        \State	time$ \leftarrow $time + $\Delta t$
        \EndFor
        \State	calculate FTLE of each particle
        \State	put the result into a new mesh using index linking

        \State	output
  \end{algorithmic}
\end{algorithm}

\subsection{The proposed refinement criterion based on modified ridge extraction method}
\label{sec:3.1}
Precious studies \cite{Garth2007Efficient,Miron2012Anisotropic,Sadlo2007Efficient} showed that the rule of where to refine the mesh of existing research remains that of traditional methods, which is according to error. This is defined in Equ.(\ref{eq:error}). This has been proved to be efficient in most cases such as \cite{Garth2007Efficient,Miron2012Anisotropic,Sadlo2007Efficient} and Sec. \ref{sec:4.1.1}. However, as in some cases , these methods may be not suitable, for the refinement particles may not be the right position.

In this article, refinement particles are placed near the FTLE ridge. Firstly, $\sigma _t^T\left( {\bf{x}} \right)$ is calculated in a coarse mesh. And then, like the method provided in \cite{Mathur2007Uncovering}, the particles for which $\left| {\nabla \sigma _t^T({\bf{x}})} \right|$ is bigger than a given threshold, and they move in the $\nabla \sigma _t^T({\bf{x}})$ field. In this paper, we define the threshold as in Equ.(\ref{eq:alpha}).
\begin{equation}\label{eq:alpha}
{\left| {\nabla \sigma _t^T({\bf{x}})} \right|_{local}} \ge \alpha {\left| {\nabla \sigma _t^T({\bf{x}})} \right|_{\max }}
\end{equation}
These particles would move until ${\nabla \sigma _t^T({\bf{x}})}$ becomes so small that the particles would hardly move.

\begin {figure}
\centering
\includegraphics[width=15cm]{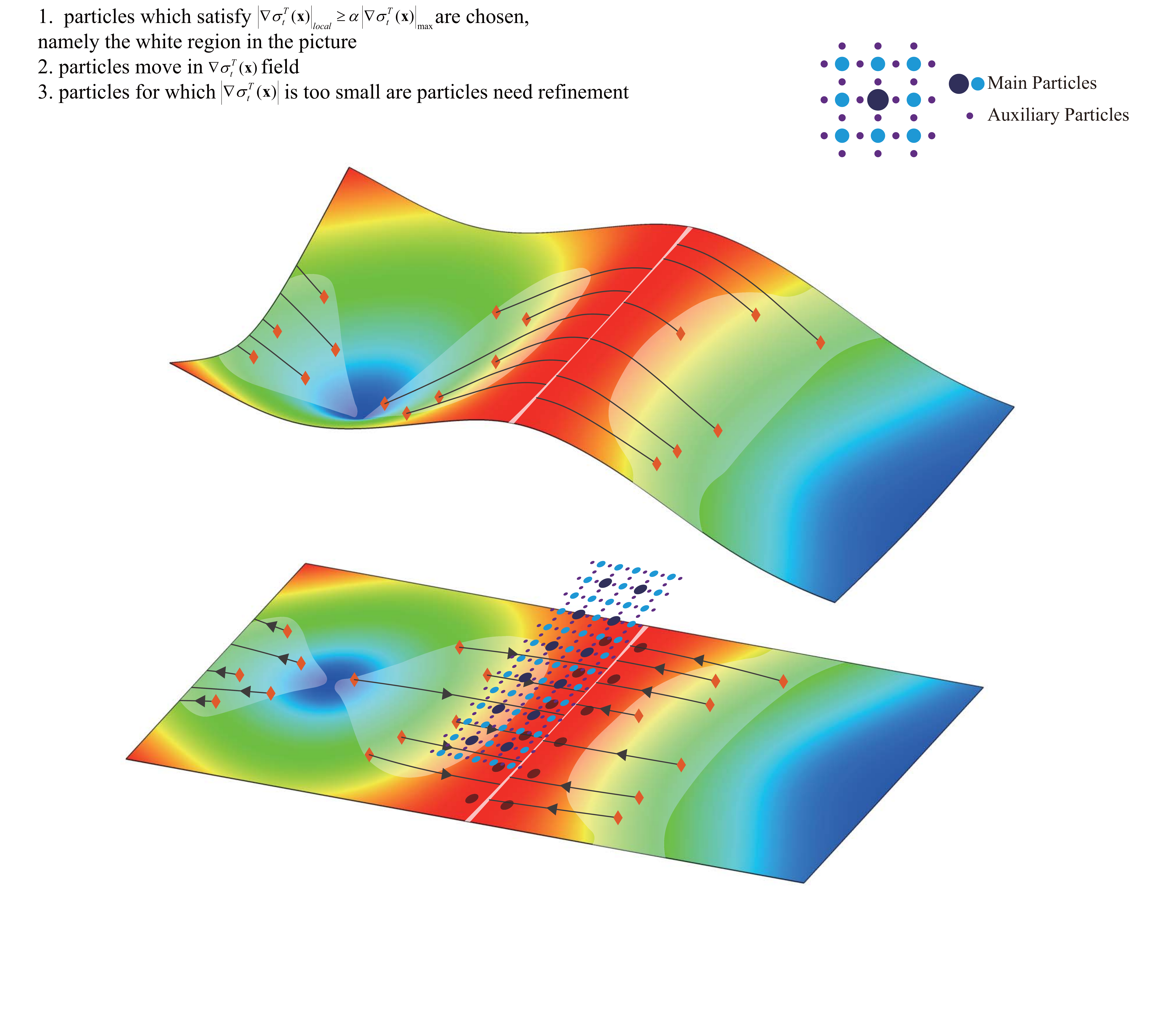}
\caption{Illustration of the objective-adaptive refinement (OAR) method}
\label{fig:illustration}
\end {figure}

When the motion of the particles are ceased, the four particles of the coarse mesh surrounding the final position are the particles which needed to be refined. When some of the particles are moving across the boundary of computational domain, the particles are seen as invalid.
The particles which are chosen firstly are not necessarily the particles need refinement, and the phenomenon is illustrated in Fig.(\ref{fig:illustration}). The firstly chosen particles are labeled as red.


For each particle which is needed for refinement, 9 Main Particles (shorten for MPs) and 24 Auxiliary Particles (shorten for APs) are placed as illustrated in the right top part of Fig.(\ref{fig:illustration}).
The MPs, namely, the blue particles in Fig.(\ref{fig:illustration}), are needed to calculate FTLE, in which the dark blue particles are the particles needed to refine the lower layer. The APs, namely the purple particles, are particles which assist in the FTLE calculation.

\subsubsection{the limiting threshold $\alpha$}

\begin {figure}
\centering
\includegraphics[width=15cm]{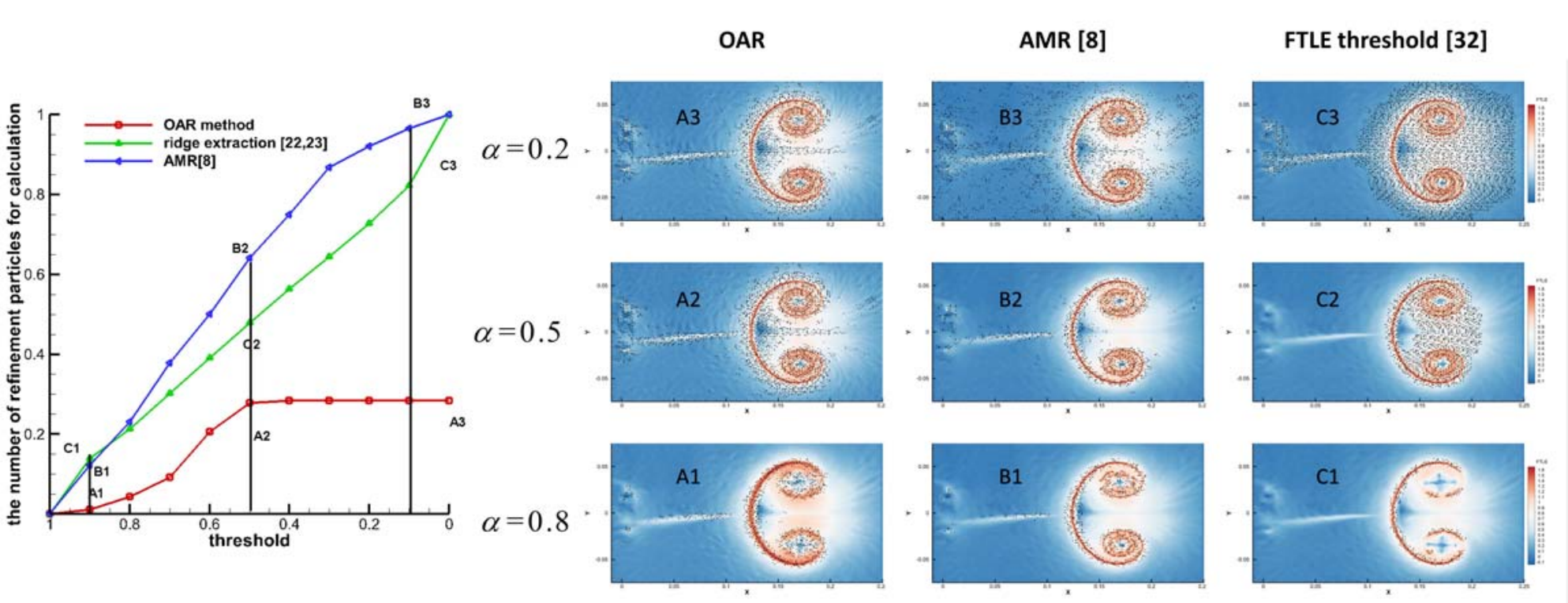}
\caption{Illustration of the influence of parameter change for different criterions}
\label{fig:alpha}
\end {figure}

It is a problem in choosing the parameter $\alpha$ mentioned in Equ.(\ref{eq:alpha}). The value of $\alpha$ does not effect to the amount of the refinement particles if it is small enough, because the particles would converge to LCS ridge though the motion, which is illustrated Fig.(\ref{fig:alpha}). We can find that compared with other criterion, the proposed refinement criterion is superior because the refinement region would not be too large to be useful when the parameter changes as showed in Fig.\ref{fig:alpha}.

For this reason, the method of finding an appropriate $\alpha$ in our algorithm is as follows. Firstly, a relatively large $\alpha$ is given, and the total number of refinement particles is counted. Then, $\alpha$ decreases by a small step $\Delta \alpha$, which is chosen to be $0.05$ in this article, and the amount of the refinement particles corresponding to the new $\alpha$ is compared with the former one. If the difference is relatively small, the $\alpha$ is the one we need. Because the refinement process includes particle motion, the calculation amount is relatively big. To reduce the calculation burden, for each $\alpha$, only the particles which haven't moved move. For the proposed method, if $\alpha$ is chosen to be too large, and thus few particles are chosen in the first step, some FTLE ridges would not be refined. On the other hand, when $\alpha$ is chosen to be small enough, the refinement region would not change.

\section{Results Comparison}
\label{sec:4}
The proposed method is compared with AMR method which is based on error \cite{Garth2007Efficient}, a ridge extraction method based on the magnitude of FTLE \cite{Sadlo2007Efficient,Lipinski2010A}, and a ridge extraction method based on gradient ascending method \cite{Mathur2007Uncovering}. Through the comparison, we can find that the proposed OAR method is superior to other methods in finding the correct refinement regions which are near the FTLE ridges.

\subsection{Bickley jet}
\label{sec:4.1.1}
The Bickley jet is an ideal model of geophysical flows \cite{Onu2014LCS}. This model is a jet like flow with counter rotating vortices, and can be an approximation of the Gulf Stream and the polar night jet perturbed by a Rossby wave. The velocity field can be given as follows:
\begin{equation}\label{eq:Bickleyjet1}
{\bf{v}}(x,y,t) = ( - {\partial _y}\psi ,{\partial _x}\psi )
\end{equation}
\begin{equation}\label{eq:Bickleyjet2}
\psi (x,y,t) = {\psi _0}(x,y) + {\psi _1}(x,y,t)
\end{equation}
\begin{equation}\label{eq:Bickleyjet3}
{\psi _0}(x,y){\rm{ = }}{{\rm{c}}_3}y - U{L_y}\tanh \frac{y}{{{L_y}}} + {\varepsilon _3}U{L_y}{{\mathop{\rm sech}\nolimits} ^2}\frac{y}{{{L_y}}}\cos {k_3}x
\end{equation}
\begin{equation}\label{eq:Bickleyjet4}
{\psi _1}(x,y,t) = U{L_y}{{\mathop{\rm sech}\nolimits} ^2}\frac{y}{{{L_y}}}\Re \left[ {\sum\limits_{n = 1}^2 {{\varepsilon _n}{f_n}(t){e^{i{k_n}x}}} } \right]
\end{equation}
As a forcing function, a solution running on the chaotic attractor of the damped and forced Duffing oscillator is chosen, specifically, as follows:
\begin{equation}\label{eq:Bickleyjet5}
\frac{{d{\phi _1}}}{{dt}} = {\phi _2}
\end{equation}
\begin{equation}\label{eq:Bickleyjet6}
\frac{{d{\phi _2}}}{{dt}} =  - 0.1{\phi _2} - \phi _1^3 + 11\cos (t)
\end{equation}
\begin{equation}\label{eq:Bickleyjet7}
{f_{1,2}}(t) = 2.625 \times {10^{ - 2}}{\phi _1}(t/6.238 \times {10^5})
\end{equation}
The parameter values are:
$U = 62.66$, ${c_2}{\rm{ = }}0.205U$, ${c_3} = 0.461U$, ${L_y} = 1.77 \times {10^6}$, ${\varepsilon _1} = 0.0075$, ${\varepsilon _2} = 0.04$, ${\varepsilon _3} = 0.3$, ${L_x} = 6.371 \times {10^6}\pi $, ${k_n} = 2n\pi /{L_x}$, ${\sigma _1} = 0.5{k_2}({c_2} - {c_3})$, ${\sigma _2} = 2{\sigma _1}$.
The integration time is $T = 4{L_x}/U$, and the domain is $x \in \left[ {0,2 \times {{10}^7}} \right]$, $y \in \left[ {{\rm{ - }}4 \times {{10}^6},4 \times {{10}^6}} \right]$.
\begin {figure}
\centering
\includegraphics[width=12cm]{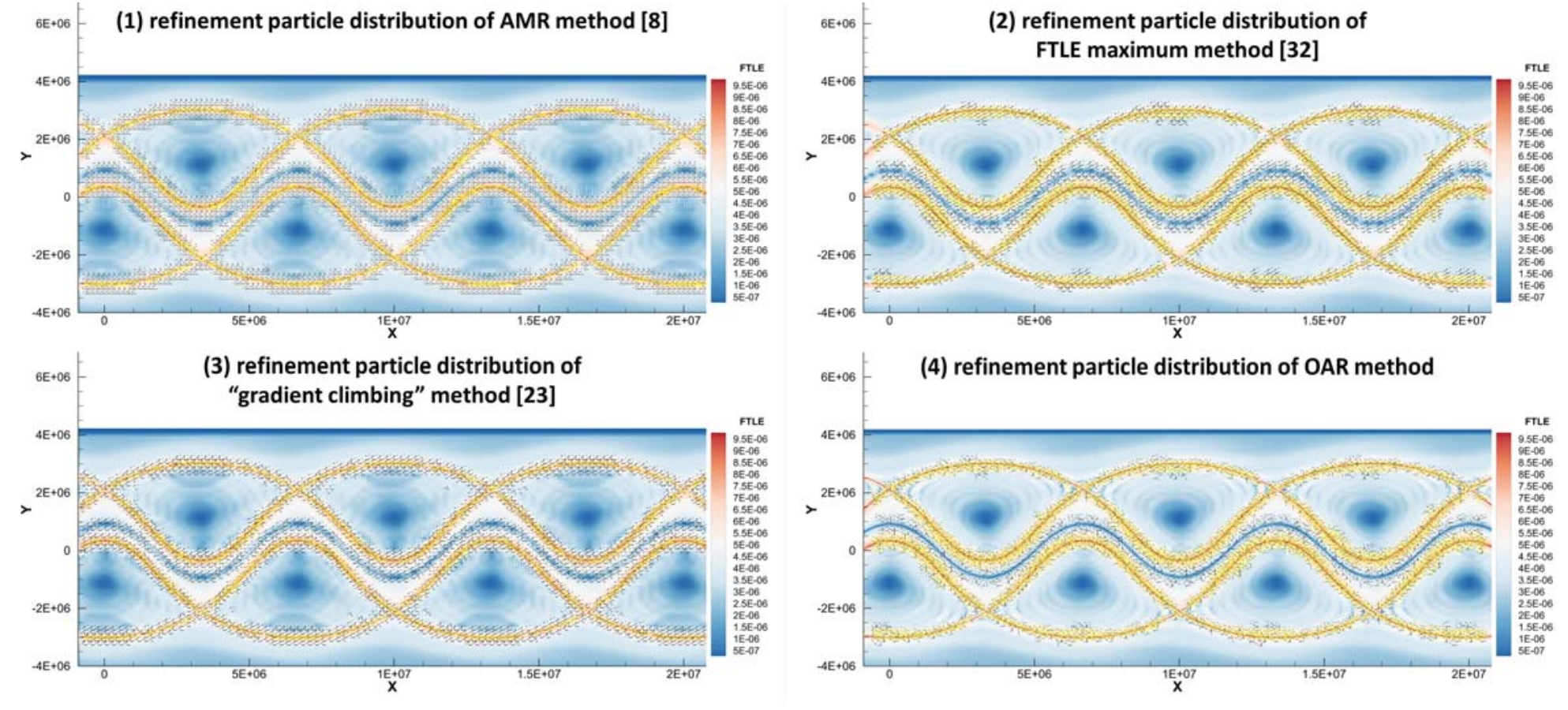}
\caption{Bickley jet model. FTLE field and refinement particle distribution using an initial mesh of $125 \times 50$ (1) AMR method \cite{Garth2007Efficient} (total particle number is $450859$) ; (b) the proposed OAR method (total particle number is $425119$) ; (c) ridge extraction method based on the magnitude of FTLE \cite{Sadlo2007Efficient,Lipinski2010A} (total particle number is $601933$) ; (d) ridge extraction method based on gradient ascending method \cite{Mathur2007Uncovering} (total particle number is $442081$)}
\label{fig:bickleyjet2}
\end {figure}
\begin {figure}
\centering
\includegraphics[width=7cm]{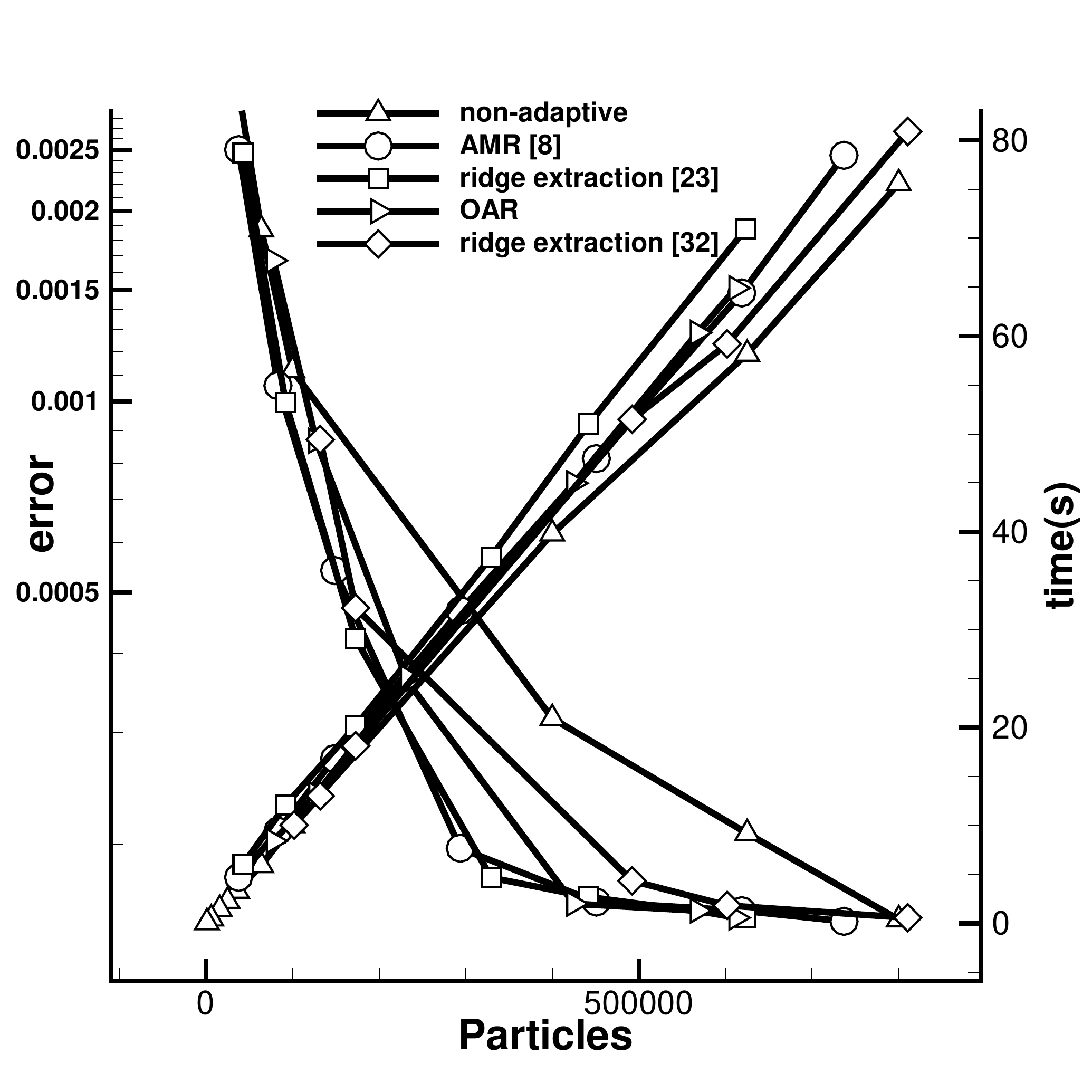}
\caption{Comparison of the errors and calculation time of the non-adaptive and different adaptive criterions}
\label{fig:bickleyjet3}
\end {figure}


Putting the results of different refinement criterions together (Fig.\ref{fig:bickleyjet2}), it can be found that the refinement regions are all near the FTLE ridge with little difference. Because of this, all these methods can reduce error efficiently, which can be illustrated in Fig.\ref{fig:bickleyjet3}.

\subsection{A case with mild FTLE ridge}
\label{sec:4.1.2}
The flow field has an interesting feature which is that there occurs an FTLE ridge at $x=0$, and the ridge is mild, which has been illustrated in \cite{Haller2011A}. The speed field is as follow
\begin{equation}\label{eq:example4}
\begin{array}{l}
\dot x = 1 + {\tanh ^2}x\\
\dot y =  - \frac{{2\tanh x}}{{{{\cosh }^2}x}}y
\end{array}
\end{equation}

For most cases, FTLE ridge is sharp, such as the case in Sec.\ref{sec:4.1.1} and Sec.\ref{sec:4.2}. However, in this case, the FTLE ridge is relatively mild, for this reason, simply applying the method in \cite{Mathur2007Uncovering} into adaptive refinement method would give false positive results and not refine the surrounding of FTLE ridge.

\begin {figure}
\centering
\includegraphics[width=9cm]{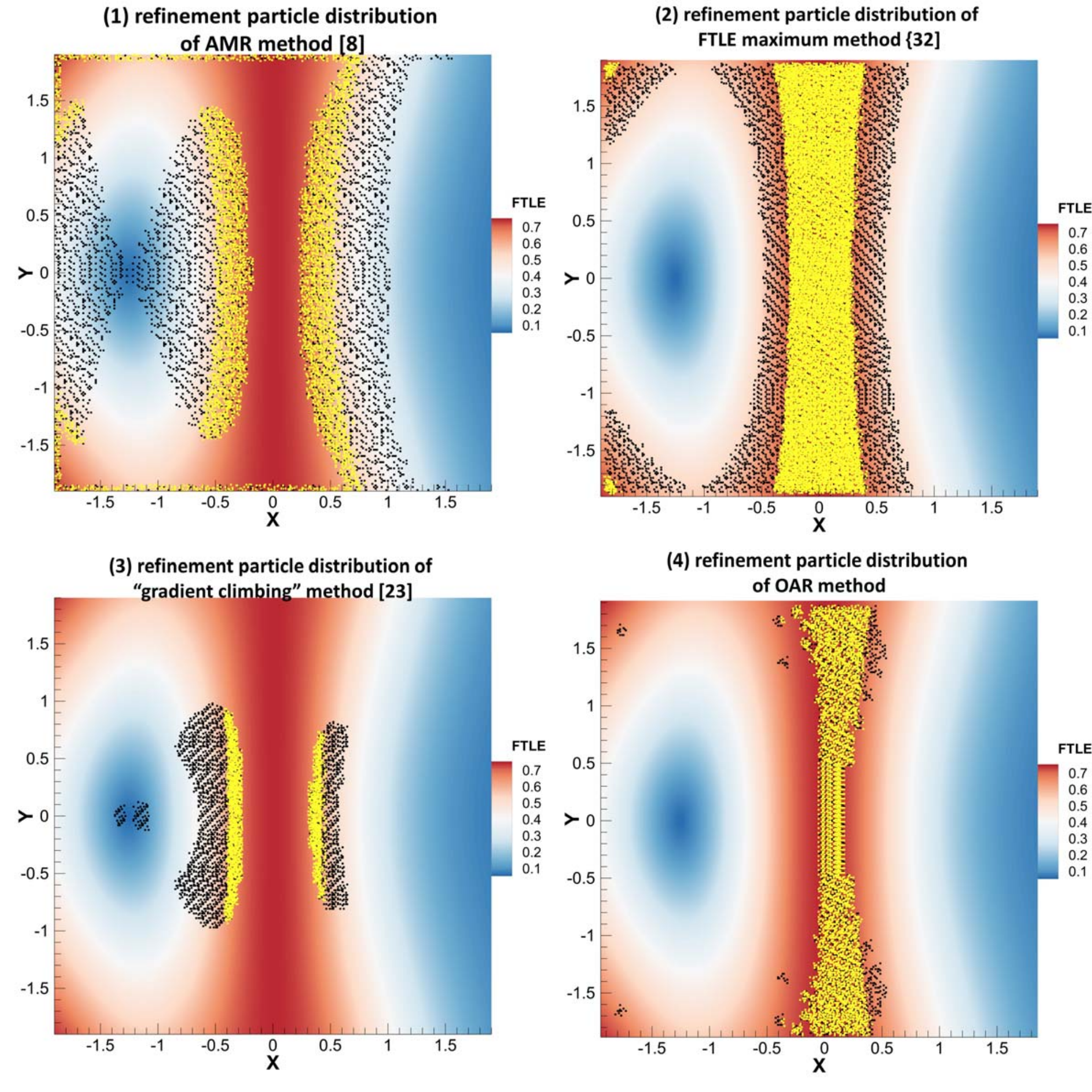}
\caption{FTLE field and refinement particle distribution using an initial mesh of $100 \times 100$ (1) AMR method \cite{Garth2007Efficient} (total particle number is $288784$) ; (b) the proposed OAR method (total particle number is $233410$) ; (c) ridge extraction method based on the magnitude of FTLE \cite{Sadlo2007Efficient,Lipinski2010A} (total particle number is $398860$) ; (d) ridge extraction method based on gradient ascending method \cite{Mathur2007Uncovering} (total particle number is $281887$)}
\label{fig:example41}
\end {figure}
The refinement particles dispersion are shown in Fig.\ref{fig:example41}. From the comparison of the four subfigures, we can found that in Fig.\ref{fig:example41}(1,4), both methods fail to refine the regions around the FTLE ridge.
\begin {figure}
\centering
\includegraphics[width=12cm]{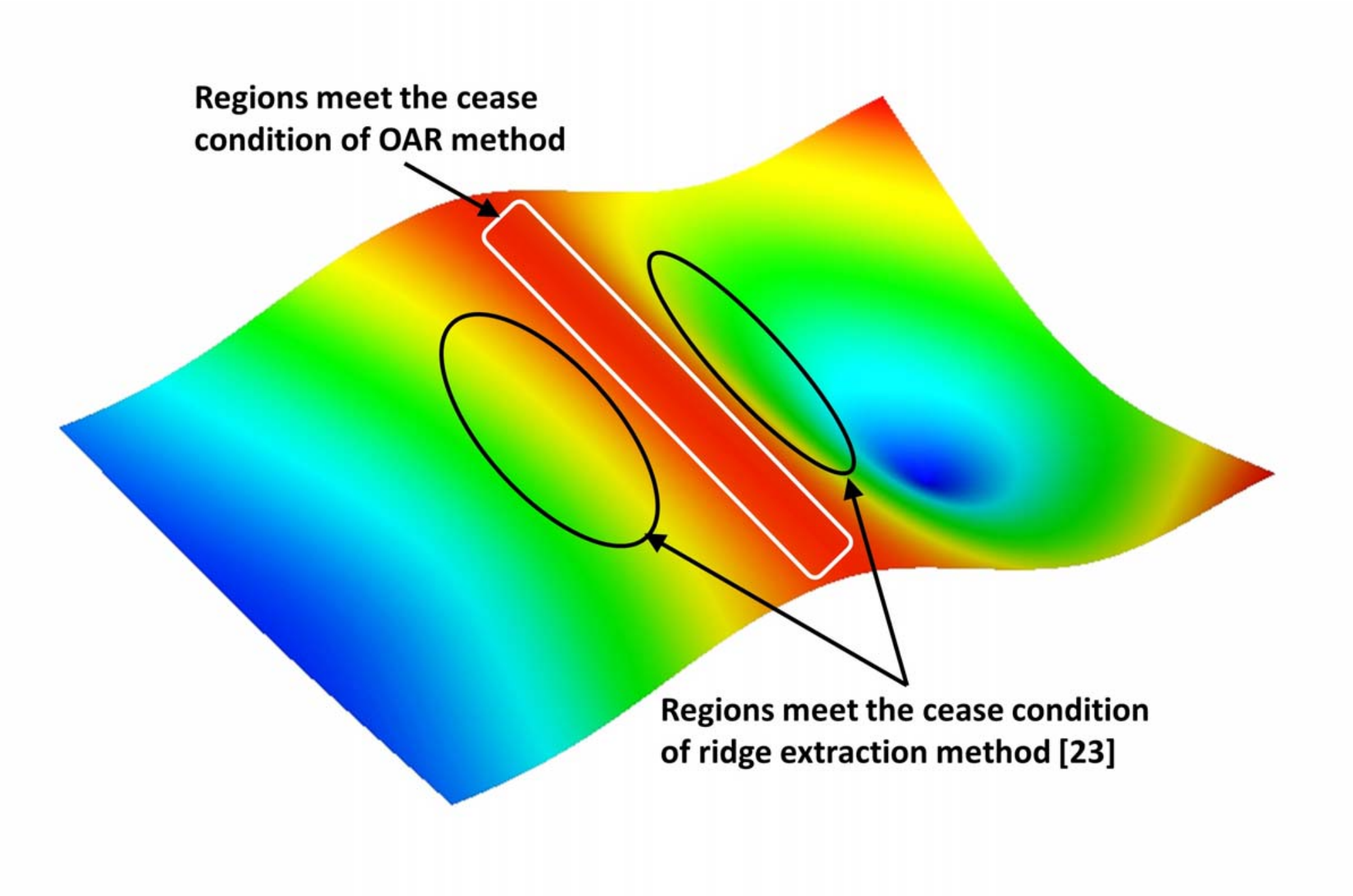}
\caption{The explanation of the refinement region of ridge extraction method based on gradient ascending method \cite{Mathur2007Uncovering}}
\label{fig:example42}
\end {figure}

\begin {figure}
\centering
\includegraphics[width=7cm]{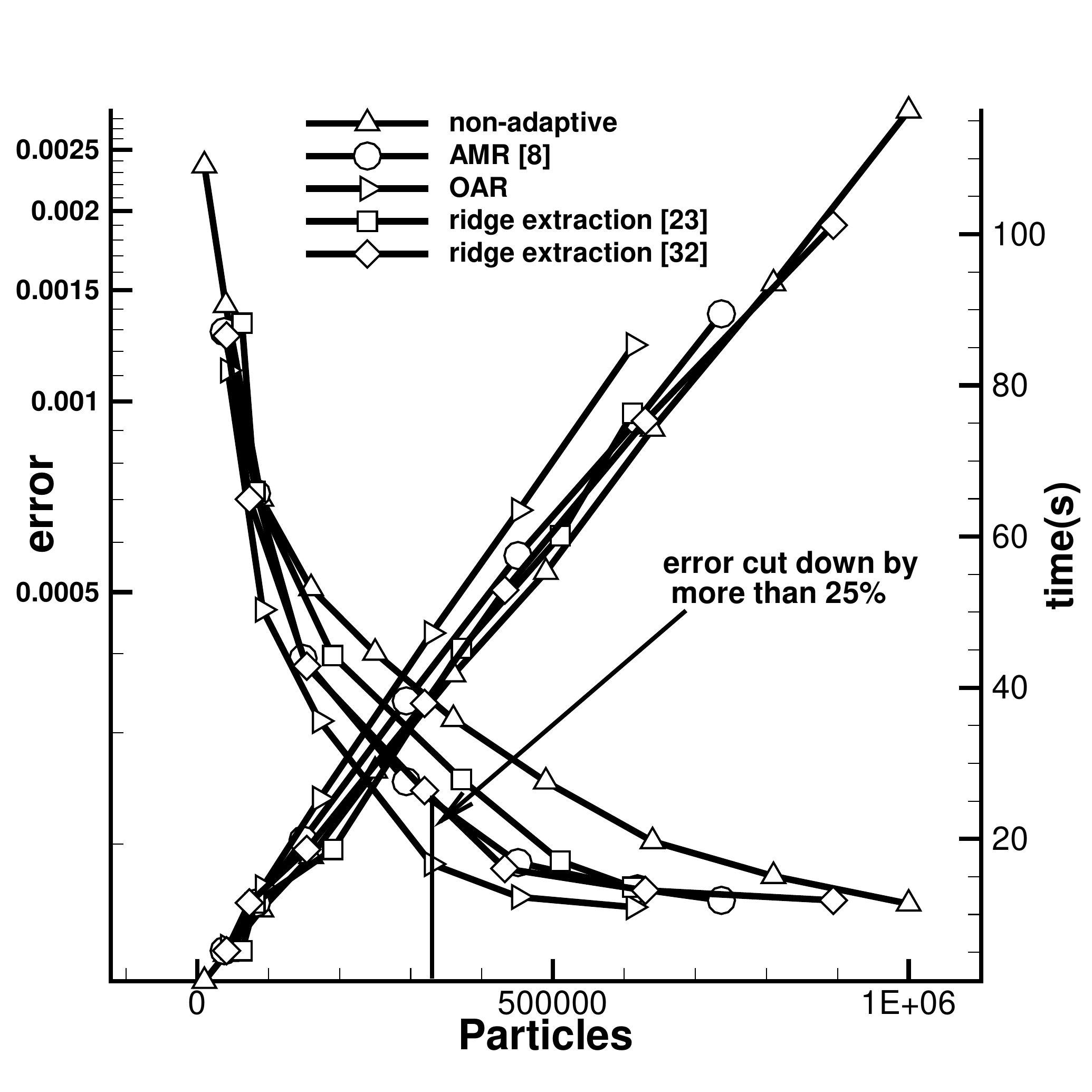}
\caption{Comparison of the errors and calculation time of the non-adaptive and different adaptive criterions}
\label{fig:example43}
\end {figure}

In our calculation, the integration time is $\Delta t = 1.78s$. From the comparison of the different methods, we can find out that using adaptive methods would cut down the error in a huge amount and slightly add to the calculation time for a definitive total particle numbers. The ridge extraction method \cite{Mathur2007Uncovering} failed to find the correct refinement region because the regions which satisfy its cease condition are on the sides on the FTLE ridge. The other ridge extraction method \cite{Sadlo2007Efficient,Lipinski2010A} though could find FTLE ridge, refines so much area that the calculation burden is very big. The proposed OAR method would spend slightly more calculation time, however it successfully find the correct refinement region, and thus cut down the error faster than other methods.

The difference in algorithm between the proposed criterion and the ridge extraction method based on gradient ascending method \cite{Mathur2007Uncovering} is the cease criterion. Fig.\ref{fig:example42} shows that the the regions that satisfy the cease criterion do not occurs to be the region near the FTLE ridge. It is because that in this case, the FTLE value near the ridge changes slowly relatively to other cases.

\subsection{Experimental cases: single vortex}
\label{sec:4.2}
LCSs have long been used in the study of the generation, formation, evolution, and pinch off of vortex rings
\cite{Shadden2006Lagrangian,O2010A,Olcay2008Measurement}. Especially in unsteady cases, LCS is superior to Euler methods in visualizing the details of vortex rings \cite{Haller2005An}.

The data we used in this article is an experimental data conducted by vortex ring generator. The experimental apparatus has been published by Suyang Qin in \cite{Qin2017Lagrangian,Qin2018On}.

\begin {figure}
\centering
\includegraphics[width=12cm]{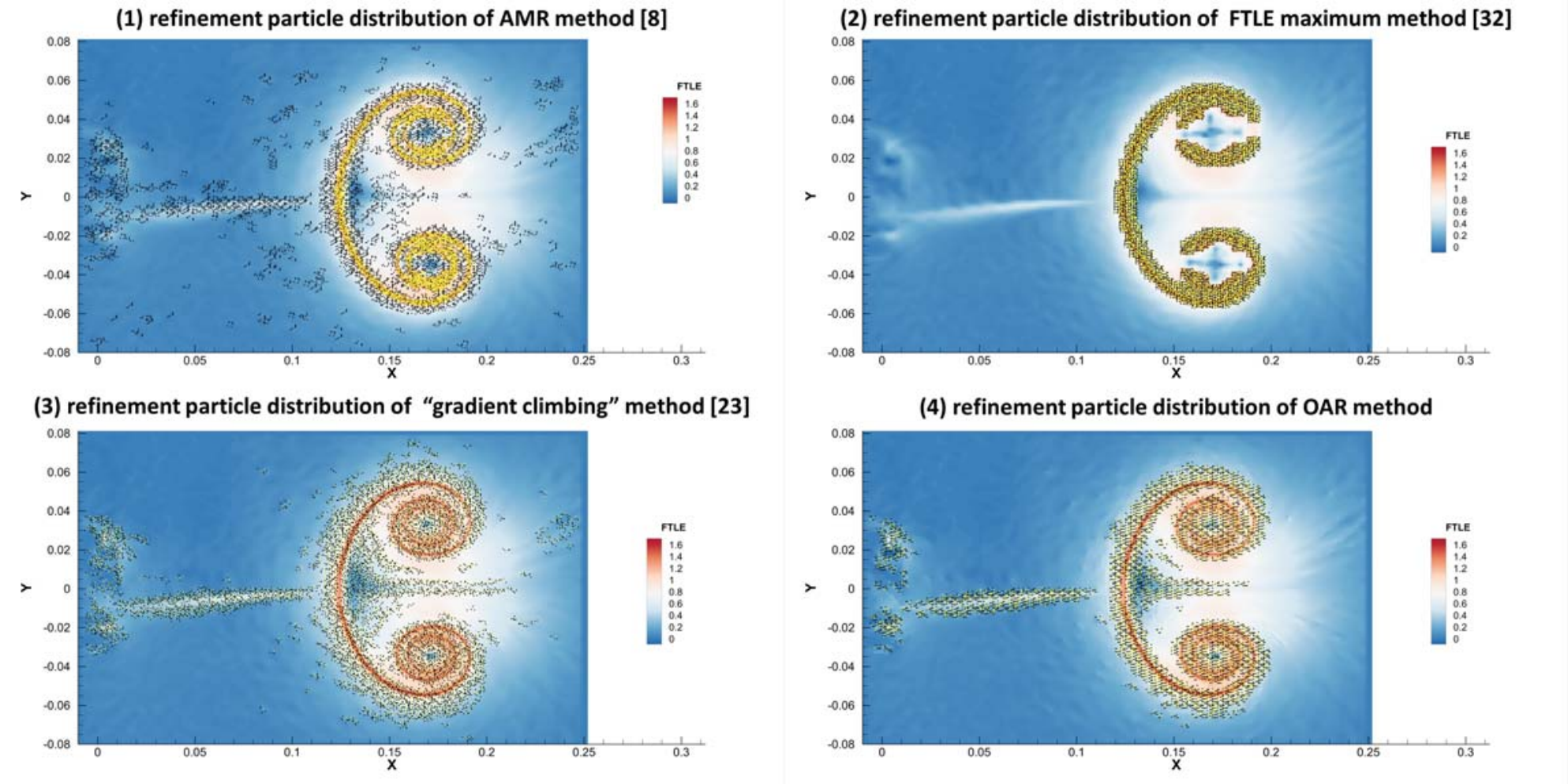}
\caption{Single vortex ring. FTLE field and refinement particle distribution using an initial mesh of $200 \times 100$ (1) AMR method \cite{Garth2007Efficient} (total particle number is $340931$) ; (b) the proposed OAR method (total particle number is $381713$) ; (c) ridge extraction method based on the magnitude of FTLE \cite{Sadlo2007Efficient,Lipinski2010A} (total particle number is $398708$) ; (d) ridge extraction method based on gradient ascending method \cite{Mathur2007Uncovering} (total particle number is $323897$)}
\label{fig:singlevortex1}
\end {figure}
\begin {figure}[H]
\centering
\includegraphics[width=7cm]{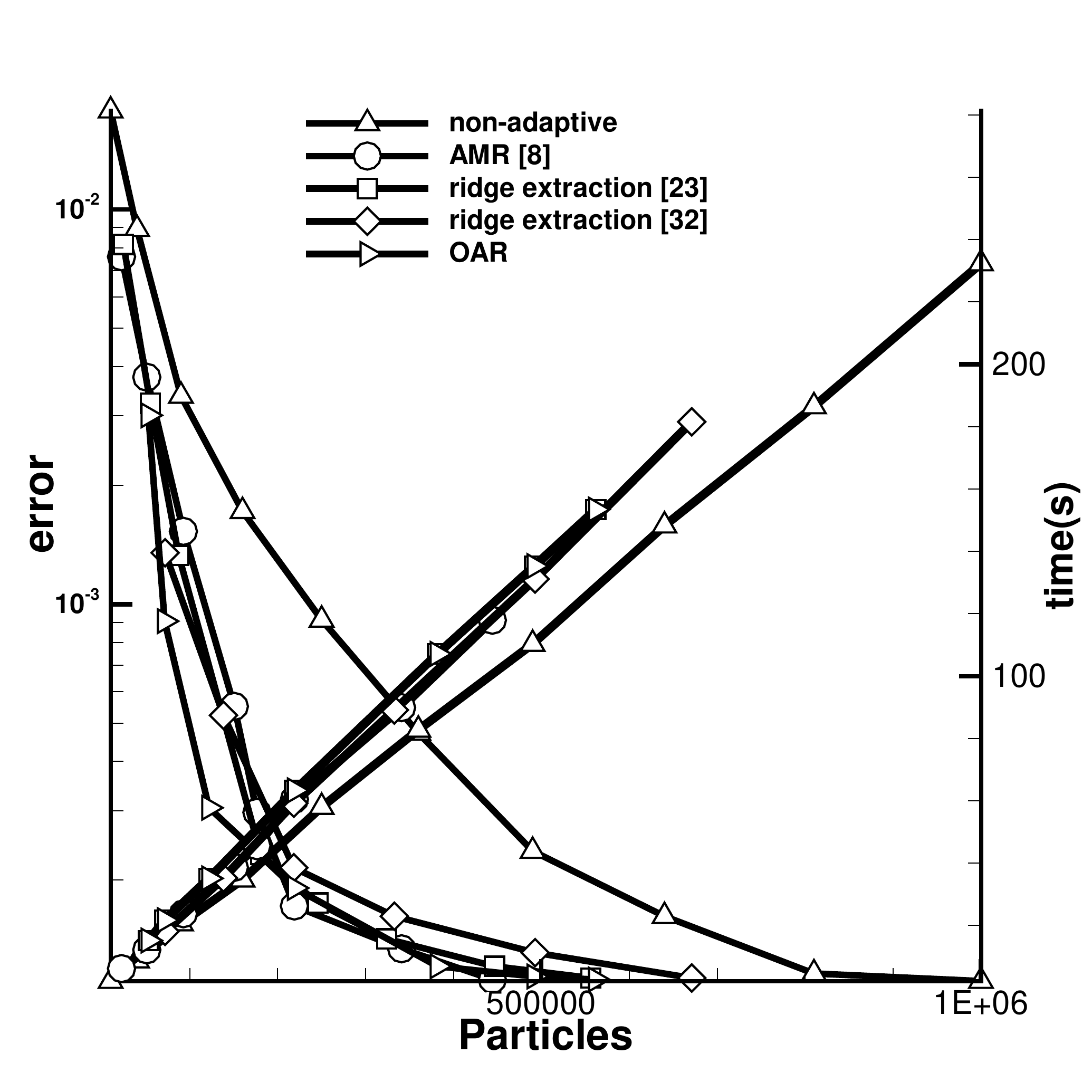}
\caption{Comparison of the errors and calculation time of the non-adaptive and different adaptive criterions}
\label{fig:singlevortex3}
\end {figure}
\begin {figure}[H]
\centering
\includegraphics[width=9cm]{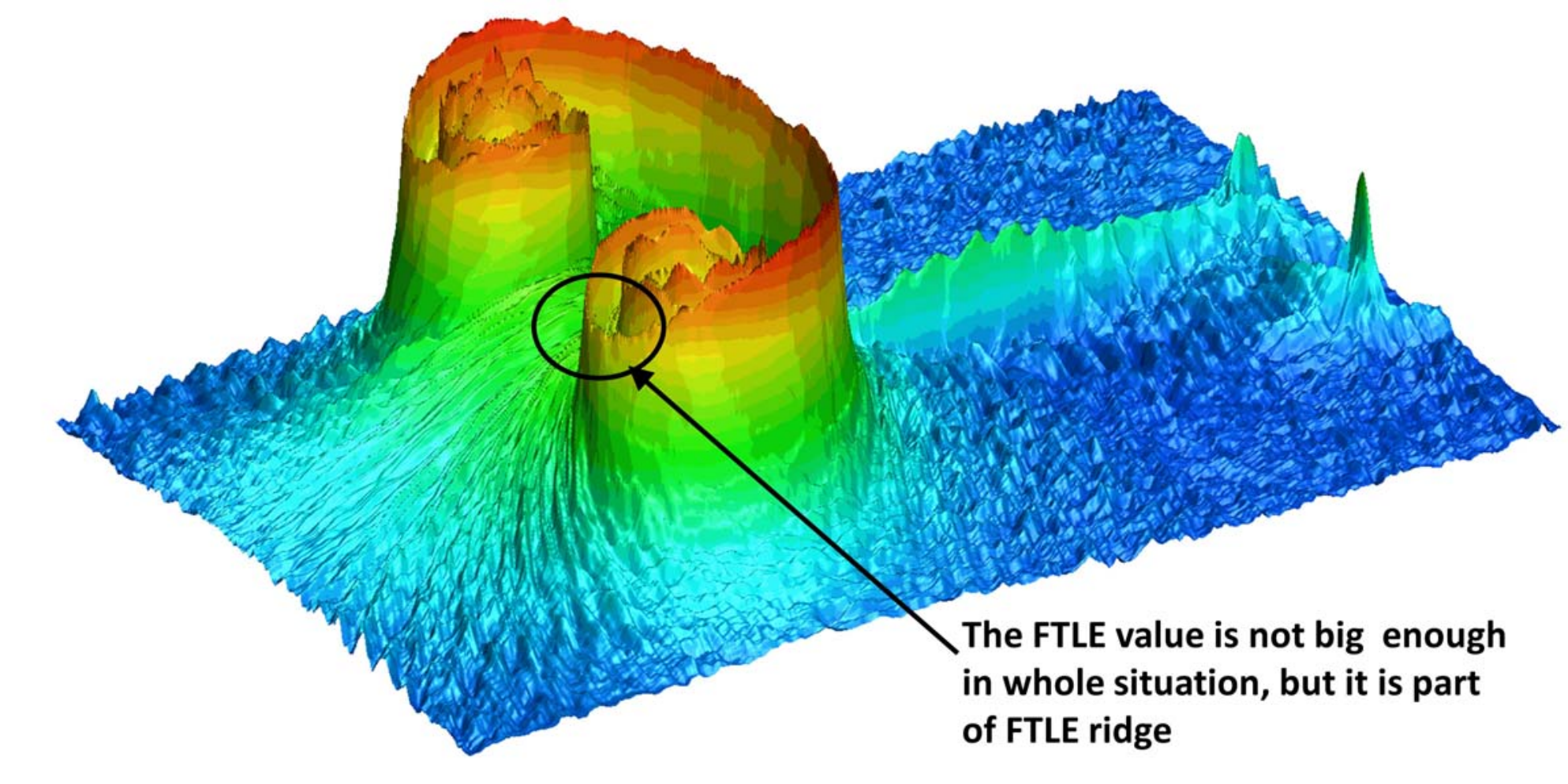}
\caption{The explanation of the refinement region of ridge extraction method based on the magnitude of FTLE \cite{Sadlo2007Efficient,Lipinski2010A}}
\label{fig:singlevortex4}
\end {figure}

Fig.\ref{fig:singlevortex1} shows vividly that the result of different refinement methods with initial coarse mesh (initial mesh of $200 \times 100$). From the comparison of the results, it can be found that the refinement of  Fig.\ref{fig:singlevortex1}(3) is nearly wrong for it fails to capture some regions of FTLE ridge. The result also shows that this refinement criterion cannot be used in cases where the experimental error in the initial data can have an influence on the FTLE field. From Fig.\ref{fig:singlevortex4}, the reason is shown that the error in initial data makes the isosurface undulate, and thus FTLE value on parts of the FTLE ridge is not enough to be big when seeing from the whole viewpoint. Because of this ,the method would give false negative result in the refinement and the result would be very bad in some certain region.
Because of the difference in the refinement regions, the ability to reduce error for different methods is different as in Fig.\ref{fig:singlevortex3}.

Fig.(\ref{fig:singlevortex1}(d)) illustrates that these characteristics are due to the higher amount of particles dispersed in the areas near the FTLE ridge. Thus, less particles are needed to achieve a given accuracy level.

The proposed OAR method can reduce error the most quickly compared with other methods, and the ridge extraction method based on the magnitude of FTLE \cite{Sadlo2007Efficient,Lipinski2010A} perfumes the worst. The quantitative results given by Fig.\ref{fig:singlevortex1} also conform with the result of refinement particle distribution. The larger the proportion of particles put near the FTLE ridge is, the faster the criterion can reduce error.

\section{Conclusions}
\label{sec:6}
In this paper, we propose the application of Objective-Adaptive Refinement (OAR) criterion based on modified ridge extraction method into finite-time Lyapunov exponent (FTLE) computation. Compared with the AMR method, the criterion has some advantages. The refinement rule can ensure that the correct position of refinement particles would be found. This character is especially important when the FTLE ridge is relatively mild or error is presented in the initial data. The efficiencies and accuracies of the proposed OAR with other adaptive refinement methods were compared by theoretical and experimental cases. Moreover, due to the characteristics of the method, the choose of the parameter is fully objective. Introducing the OAR criterion into LCS computation results in the improvement in the calculation efficiency and calculation accuracy. The proposed algorithm can be introduced in the first step of calculating new LCSs, and can also be extended to three-dimensional cases, which are the future focuses of this study.

\begin{acknowledgements}
The authors would like to thank the Center for High Performance Computing of SJTU for providing the super computer $\pi$ to support this research. This work is supported by the National Natural Science Foundation of China (NSFC-91441205, NSFC-91741113) and National Science Foundation for Young Scientists of China (Grant No. 51606120). Furthermore, Chunhui Tang, Haiyan Lin, and Geng Liang are appreciated in assisting the completion of the paper.
\end{acknowledgements}


\bibliography{reference}   
%

\end{document}